\documentclass[twocolumn,superscriptaddress,reprint]{revtex4-1}

\usepackage{amsmath}
\usepackage{amssymb}
\usepackage{graphicx}
\usepackage{color}  

\usepackage[dvipdfm]{hyperref} 

\newcommand{\qw} {(\mathbf{q},\omega)}

\begin{document}

\title{The spin anisotropy of the magnetic excitations in the normal and superconducting states of optimally doped YBa$_{2}$Cu$_{3}$O$_{6.9}$ studied by polarized neutron spectroscopy}

\author{N. S. Headings}
\affiliation{H.H. Wills Physics Laboratory, University of Bristol,
Tyndall Ave., Bristol, BS8 1TL, UK}
\author{S. M. Hayden}
\email{s.hayden@bris.ac.uk} \affiliation{H.H. Wills Physics
Laboratory, University of Bristol, Tyndall Ave., Bristol, BS8 1TL,
UK}
\author{J. Kulda}
\affiliation{Institut Laue-Langevin, BP 156, 38042 Grenoble, France}
\author{N. Hari Babu}
\affiliation{Engineering Department, Trumpington Street, University
of Cambridge, CB2 1PZ, UK}
\author{D. A. Cardwell}
\affiliation{Engineering Department, Trumpington Street, University
of Cambridge, CB2 1PZ, UK}

\pacs{74.72.Gh,75.40.Gb,78.70.Nx,71.45.Gm}
\begin{abstract}
We use inelastic neutron scattering with spin polarization analysis
to study the magnetic excitations in the normal and superconducting
states of YBa$_{2}$Cu$_{3}$O$_{6.9}$. Polarization analysis allows
us to determine the spin polarization of the magnetic excitations
and to separate them from phonon scattering. In the normal state, we
find unambiguous evidence of magnetic excitations over the
10--60~meV range of the experiment with little polarization dependence to the excitations. In the superconducting state, 
the magnetic response is enhanced near the ``resonance energy'' and above. At lower energies, 10$\lesssim$$E$$\lesssim$30~meV, the local susceptibility becomes anisotropic, with the excitations polarized along the c-axis being suppressed.   
We find evidence for a new diffuse anisotropic response polarized perpendicular to the $c$-axis which may carry significant
spectral weight.
\end{abstract}

\maketitle

\section{Introduction}

High temperature superconductivity (HTS) arises when certain two
dimensional antiferromagnetic Mott insulators are electron or hole
doped \cite{HTCReviews}.  The antiferromagnetic parent compounds
such as La$_2$CuO$_4$ show spin-wave excitations up to
$2J$$\approx$300~meV \cite{Headings2010a}. Doping causes
the magnetic response to evolve from that of spin waves to a more
structured response
\cite{Cheong1991a,Rossat1991a,Mook1993a,Fong1995a,Bourges1997a,Mook1998a,Dai1999a,Hayden2004a,Vignolle2007a,Fauque2007a,Xu2009a},
with strong spin fluctuations being observed for superconducting
compositions in a number of systems including YBa$_2$Cu$_3$O$_{6+x}$
(YBCO)
\cite{Mook1998a,Bourges1997a,Dai1999a,Hayden2004a,Stock2005a,Woo2006a},
La$_{2-x}$Sr$_{x}$CuO$_4$ \cite{Hayden1996a,Vignolle2007a} and
Bi$_2$Sr$_2$CaCu$_2$O$_{8+\delta}$
\cite{Fong1999a,Fauque2007a,Xu2009a}.  Many optimally-doped cuprates
show a strong well-defined collective magnetic excitation which is
localised in reciprocal space and strongest near the
$\mathbf{Q}$=(1/2,1/2)$\equiv$($\pi,\pi$) position. It  is sharp in
energy and develops on cooling through the critical temperature.
This excitation has become known as the ``magnetic resonance''. The
magnetic resonance has been observed in YBa$_2$Cu$_3$O$_{6+x}$
\cite{Rossat1991a,Mook1993a,Fong1995a},
Bi$_2$Sr$_2$CaCu$_2$O$_{8+\delta}$ \cite{Fong1999a},
Tl$_2$Ba$_2$CuO$_{6+\delta}$ \cite{He2002a} and
HgBa$_2$CuO$_{4+\delta}$ \cite{Yu2010a}.

The magnetic resonance is certainly the strongest feature in the
magnetic excitations spectrum of the materials listed above,
however, it only accounts for a small faction ($\approx$2\%) \cite{Dai1999a,Fong1999a,Woo2006a} of the
total scattering expected from the unpaired 3$d$ electrons of the Cu
atoms. In this work we search for other contributions to the
response which are spread out in energy and wavevector but
nevertheless may carry significant spectral weight. These are harder
to observe because they are weak and may not
show the strong temperature dependence which allows the resonance to be easily
isolated. We use inelastic neutron scattering with polarization
analysis to isolate the magnetic scattering from phonon scattering.

We find that there is a significant response in the normal state
which can account for much of the spectral weight from which the
resonance is formed. In the superconducting state, we find evidence
for a diffuse contribution at energies well below the resonance.
This new contribution is polarized with strong fluctuations perpendicular to 
the $c$-axis.

\section{Background}

\subsection{Polarization Analysis}
Neutrons scatter from condensed matter via two processes: (i) The
electromagnetic interaction probes fluctuations in the magnetization
density of the electrons (in this paper this is referred to as
magnetic scattering). (ii) The strong nuclear force is responsible
for scattering from the atomic nuclei. The nuclear scattering allows
us to probe phonons which are correlations (in time and space)
between the position of the nuclei. The existence of two distinct
scattering processes makes the neutron an extremely versatile probe.
However, it also means that the two types of scattering can mask
each other.

Polarization analysis of the neutron's spin allows the separation of
magnetic and nuclear (phonon) scattering.   In the present work, we use
longitudinal polarization analysis (LPA). In LPA, a spin-polarized
incident neutron beam is created and its polarization maintained by
a small magnetic field ($\sim$1~mT).  The number of neutrons
scattered with spins parallel or antiparallel to this quantizing
field are then measured.  We label each spin-polarization state as
parallel ($\equiv$up,$\uparrow$,+) or antiparallel
($\equiv$down,$\downarrow$,$-$) to the applied field.  The cross
sections are referred to as spin-flip (SF)
($\uparrow\rightarrow\downarrow$,$\downarrow\rightarrow\uparrow$) or
non-spin-flip (NSF)
($\uparrow\rightarrow\uparrow$,$\downarrow\rightarrow\downarrow$).
A natural reference frame in which to
understand the cross sections is one referenced to
the scattering vector $\mathbf{Q}=\mathbf{k}_{i}-\mathbf{k}_{f}$ of
the neutron, where $\mathbf{k}_{i}$ and $\mathbf{k}_{f}$ are the
incident and final wavevectors of the neutron. Thus,
$\mathbf{\hat{x}} \parallel \mathbf{Q}$, $\mathbf{\hat{y}} \perp
\mathbf{Q}$, and $\mathbf{\hat{z}} \perp \mathbf{Q}$ and $\perp$ to
the spectrometer scattering plane (the plane containing
$\mathbf{k}_{i}$ and $\mathbf{k}_{f}$). We make measurements with
the neutrons polarized along each of these axes.

The neutron cross sections as a function of spin polarization have
been derived and presented elsewhere
\cite{Blume1963a,Maleyev1963a,Moon1969a,Squires1978a,Lorenzo2007a}.
The spin-flip \textit{magnetic} cross section for spin polarization
$\parallel \mathbf{Q}$ is
\begin{eqnarray}
\nonumber
  \sigma_{xx}^{\uparrow \downarrow} = \left( \frac{d^2 \sigma}{d \Omega dE} \right)_{\mathbf{H} \parallel x}^{\uparrow \rightarrow \downarrow}  &=& \frac{k_f}{k_i} \frac{(\gamma r_e)^2}{g^2 \mu^2_B} \frac{1}{\pi} F^2(\mathbf{Q}) \\
  & \times &  \frac{\chi_{yy}^{\prime\prime}({\bf q},\hbar\omega)+\chi_{zz}^{\prime\prime}({\bf q},\hbar\omega)}{1-\exp(-\hbar\omega/kT)},
  \label{Eqn:cross_section}
\end{eqnarray}
where $(\gamma r_{\text{e}})^2$=0.2905 barn sr$^{-1}$ and $|F({\bf
Q})|^2$ is the anisotropic magnetic form factor for a Cu$^{2+}$
$d_{x^{\scriptstyle 2}-y^{\scriptstyle 2}}$ orbital.
$\chi_{\nu\nu}^{\prime\prime}({\bf q},\hbar\omega)$ is the
generalized susceptibility corresponding to magnetic fluctuations
along the $\nu$-axis. Thus, for example:
\begin{equation}
\langle m_x^2({\bf q},\omega) \rangle = \frac{1}{\pi}
\frac{\chi_{xx}^{\prime\prime}({\bf
q},\omega)}{1-\exp(-\hbar\omega/kT)}, \label{Eqn:FD_theorem}
\end{equation}
where the angle brackets denote thermal averages.  The
spin-dependent cross sections including the nuclear coherent cross
sections (i.e. the phonon cross section) $N \qw$  are:
\begin{eqnarray}
\sigma_{xx}^{\uparrow \downarrow} &\propto& \chi_{yy}^{\prime\prime} \qw + \chi_{zz}^{\prime\prime} \qw    + \mathrm{BG}_{\uparrow \downarrow} \nonumber \\
\sigma_{yy}^{\uparrow \downarrow} &\propto&                                \chi_{zz}^{\prime\prime} \qw    + \mathrm{BG}_{\uparrow \downarrow} \nonumber \\
\sigma_{zz}^{\uparrow \downarrow} &\propto& \chi_{yy}^{\prime\prime} \qw                                   + \mathrm{BG}_{\uparrow \downarrow} \nonumber \\
\sigma_{xx}^{\uparrow \uparrow}   &\propto& N \qw                                                          + \mathrm{BG}_{\uparrow \uparrow} \nonumber \\
\sigma_{yy}^{\uparrow \uparrow}   &\propto& \chi_{yy}^{\prime\prime} \qw  + N \qw                          + \mathrm{BG}_{\uparrow \uparrow} \nonumber \\
\sigma_{zz}^{\uparrow \uparrow}   &\propto&
\chi_{zz}^{\prime\prime}\qw + N \qw +
\mathrm{BG}_{\uparrow\uparrow},
\label{Eqn:all_sigmas}
\end{eqnarray}
where we have neglected the nuclear spin incoherent cross-section
which is small in the present experiments \cite{Cross_sec_note} and
BG denotes the background for the configuration. In this work we
isolate two components of the susceptibility by comparing different
SF cross sections:
\begin{eqnarray}
\label{Eq:sigma_diff}
\sigma_{xx}^{\uparrow \downarrow} - \sigma_{yy}^{\uparrow \downarrow} &\propto& \chi_{yy}^{\prime\prime} \qw \nonumber \\
\sigma_{xx}^{\uparrow \downarrow} - \sigma_{zz}^{\uparrow
\downarrow} &\propto& \chi_{zz}^{\prime\prime} \qw
\end{eqnarray}

\subsection{Bilayer Effects}
\begin{figure}
\begin{center}
\includegraphics[width=0.4\linewidth]{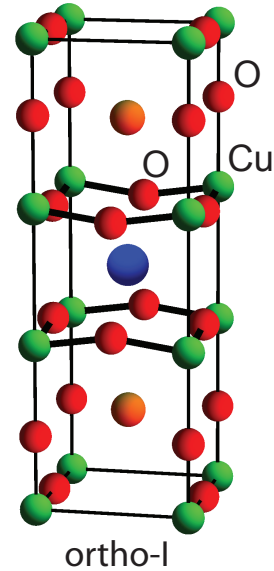}
\end{center}
\caption{(Color online) The ortho-I structure of
YBa$_{2}$Cu$_{3}$O$_{6.9}$.} \label{Fig:YBCO_struct}
\end{figure}
YBa$_{2}$Cu$_{3}$O$_{6+x}$ has two CuO$_2$ planes per unit cell (See
Fig.~\ref{Fig:YBCO_struct}). The usual starting point for models of
the magnetic response is to neglect the electronic coupling between
CuO$_2$ planes in different unit cells and include only coupling
between the CuO$_2$ planes of the bilayer located in the center of the unit
cell in Fig.~\ref{Fig:YBCO_struct}. This leads to a pair of bonding
(b) and antibonding (a) energy bands. The presence of a mirror plane
between the two planes of the bilayer means that the magnetic
excitations have distinct odd (o) or even (e) character.  In this description, the
magnetic response is of the form
\cite{Bulut1996a,Millis1996a,Brinckmann2001a,Eremin2007a}
\begin{eqnarray}
\nonumber \chi^{\prime\prime}(h,k,l,\omega) &=&
\chi_{e}^{\prime\prime}(h,k,\omega) \cos^2 \left( \frac{\pi d }{c} l \right) \\
&+& \chi_{o}^{\prime\prime}(h,k,\omega) \sin^2 \left( \frac{\pi d
}{c}  l \right), \label{Eqn:chi_oe_def}
\end{eqnarray}
where $d$ is the separation of the CuO$_2$ planes. For
YBa$_{2}$Cu$_{3}$O$_{6.9}$ $d$=3.38~\AA, this means the odd response
is strongest at $l$$=$$(n+1/2)c/(2d)$=1.73, 5.3,$\ldots$ The
strongest features in the magnetic response of
YBa$_{2}$Cu$_{3}$O$_{6+x}$ observed by INS are in the odd channel
\cite{Rossat1991a,Mook1993a,Fong1995a} and we measure the odd
channel in the present experiment. We note that weaker resonance
features have been reported in the even channel
\cite{Pailhes2004a,Pailhes2006a} for various dopings. The reported
even resonance occurs at higher energy than in the odd channel.

\subsection{Sample Details}
We investigated a near optimally doped sample of
YBa$_{2}$Cu$_{3}$O$_{6.9}$ ($T_c$=93~K) grown by a top seed melt
growth technique \cite{Babu2000a}. YBa$_{2}$Cu$_{3}$O$_{6.9}$ has
the ortho-I structure show in Fig.~\ref{Fig:YBCO_struct} with
lattice parameters $a$=3.82~\AA, $b$=3.89~\AA\ and $c$=11.68~\AA\
($T$=77~K) \cite{Jorgensen1990a}. The single crystal studied in the
present experiment is twinned and the results presented are an
average over the two twin domains. The crystal had a mass of 32.5~g
and mosaic spread 1.3$^{\circ}$. It was annealed for 17 days at
550$^{\circ}$C, followed by 13 days at 525$^{\circ}\mathrm{C}$, in
oxygen to achieve the required oxygen stoichiometry. Neutron
depolarization measurements (see Fig.~\ref{Fig:YBCO9_q_raw})
indicated that $T_c (\mathrm{onset})=93$$\pm$0.2~K.  Based on $T_c$
and the heat treatment \cite{Meuffels1989a,Liang2006a}, we estimate
the oxygen stoichiometry to be $x$=0.9$\pm$0.01.

\subsection{Experimental Setup}
\begin{figure}
\begin{center}
\includegraphics[width=0.8\linewidth]{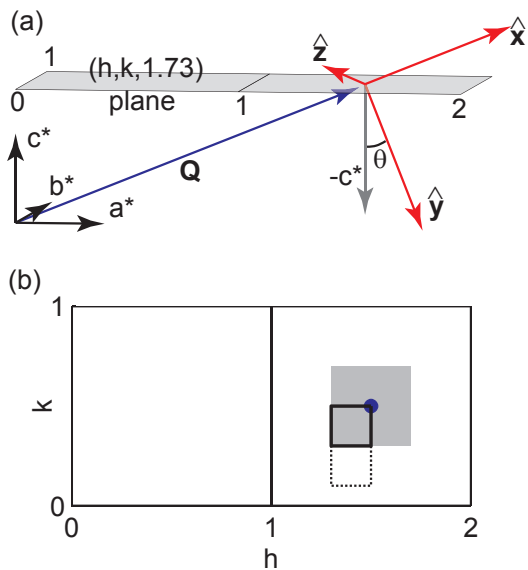}
\end{center}
\caption{(Color online)(a) Illustration of the reference frame used
to describe polarization analysis. Neutrons are polarized along the
$\hat{\mathbf{x}}$, $\hat{\mathbf{y}}$ or $\hat{\mathbf{z}}$  axes.
$\hat{\mathbf{x}}$ is parallel to $\mathbf{Q}$, $\hat{\mathbf{z}}$
is perpendicular to $\hat{\mathbf{x}}$ and in the $(h,k,0)$ plane.
Thus $\hat{\mathbf{x}} \parallel (1.5 \mathbf{a}^{\star}$+$0.5
\mathbf{b}^{\star}$+$1.8 \mathbf{c}^{\star})$, $\hat{\mathbf{y}}
\parallel (0.54 \mathbf{a}^{\star}$+$0.18 \mathbf{b}^{\star}$$-$$4.6
\mathbf{c}^{\star})$, $\hat{\mathbf{z}} \parallel (-0.5
\mathbf{a}^{\star}$+$1.5 \mathbf{b}^{\star})$, $\theta=20.6^{\circ}$
and $\cos^2 \theta=0.88$. (b) Illustration of the area in reciprocal space where
the measurements in Sec.~\ref{Sec:local} were made. For $E < 52$~meV,
we used data collected over the black square (1.3$\leq$$h$$\leq$1.5 and
0.3$\leq$$k$$\leq$0.5) to infer $\chi^{\prime\prime}(\omega)$
measured over the grey area. Data in Fig.~\ref{Fig:YBCO9_maps}
covers the black square plus dotted area.} \label{Fig:geo_fig.eps}
\end{figure}

Experiments were performed using the IN20 three-axis spectrometer at
the Institut Laue-Langevin, Grenoble using a standard longitudinal
polarization analysis set up. Neutron polarization analysis was
carried out using a focussing Heusler monochromator and analyzer.
The sample was mounted with the [310] and [001]
directions in the horizontal scattering plane of the instrument. We
worked around the (1.5,0.5,1.73) reciprocal space position so as to
avoid strong phonon scattering near $E \approx$40~meV
\cite{Fong1995a}. We used a sample goniometer to access reciprocal
space positions out of the $(3h,h,l)$ plane. Data were converted to
an absolute scale using a vanadium standard and
Eq.~\ref{Eqn:cross_section} and measurement of an acoustic phonon at
$\mathbf{Q}$=(0.2,0.2,6). The overall error in the absolute scale is about 20\%.
We use the reciprocal space of the
average tetragonal lattice (with $a \approx 3.86$~\AA) to label
wavevectors with $\mathbf{Q}=h \mathbf{a}^{\star}+ k
\mathbf{b}^{\star}+l \mathbf{c}^{\star}$.

In order to reduce neutron depolarization for measurements made in
the superconducting state, the sample was cooled through $T_c$ and
to 10~K while shielded by a $\mu$-metal shield such that $\mu_0
H<$0.3~$\mu$T. During the measurement, fields in the range $|\mu_0
\mathbf{H}|$ =0.7-0.11~mT $\ll$ $\mu_0 H_{c1}$$\approx$25-85~mT were
applied to the sample. Therefore, the sample was in the Meissner state.

The finite polarization of the incident neutron beam and other
instrumental imperfections leads to a mixing of the spin-flip and
non spin-flip channels.  This can be
described by a flipping ratio $F$, where the measured cross section
is:
\begin{equation}
\sigma_{\mathrm{SF}}^{\mathrm{meas}} = \left( \frac{F}{F+1} \right)
\sigma_{\mathrm{SF}} + \left( \frac{1}{F+1} \right)
\sigma_{\mathrm{NSF}}.
\end{equation}
We corrected our data for this mixing using the standard equations
\cite{Lipscombe2010a}:
\begin{equation}
\sigma_{\mathrm{SF}}^{\mathrm{corr}} = \left( \frac{F}{F-1} \right)
\sigma_{\mathrm{SF}}^{\mathrm{meas}} - \left( \frac{1}{F-1} \right)
\sigma_{\mathrm{NSF}}^{\mathrm{meas}},
\end{equation}
where the flipping ratio $F \approx 7.5$ was determined from
measurements on Bragg peaks made under the same conditions. For
experimental reasons, measurements were made with neutrons polarized
parallel and perpendicular to the scattering vector $\mathbf{Q}$
which meant that the neutron polarizations and hence the measured
susceptibilities are not along the crystallographic axes (see
Fig.~\ref{Fig:geo_fig.eps}). For example, the angle between the
$y$-axis and the crystallographic $c$-axis is
$\theta$=20.6$^{\circ}$. This leads to a small mixing of the
different components of the susceptibility during the measurement.
Thus:
\begin{eqnarray}
\nonumber \sigma_{xx}^{\uparrow\downarrow} - \sigma_{yy}^{\uparrow\downarrow} & \propto & 0.11\chi_a^{\prime\prime} +0.01\chi_b^{\prime\prime}+0.88\chi_c^{\prime\prime} (\equiv \chi_{c}^{\prime\prime}) \\
\sigma_{xx}^{\uparrow\downarrow} -
\sigma_{zz}^{\uparrow\downarrow} & \propto &
0.1\chi_a^{\prime\prime} +0.9\chi_b^{\prime\prime} \ (\equiv
\chi_{a/b}^{\prime\prime}).
\label{Eqn:sigma2abc}
\end{eqnarray}
This mixing does not affect the conclusions of the paper and we have
not corrected for it. We refer to the two components above as
$\chi_{a/b}^{\prime\prime}$ and $\chi_{c}^{\prime\prime}$. The local
susceptibility (see Sec \ref{Sec:chi_local}) was estimated by
measuring a grid of 36 points over the area $1.3 \leq h \leq 1.5$
and $0.3 \leq k \leq 0.5$ (at the highest energy we used $1.5 \leq h
\leq 1.7$ and $0.5 \leq k \leq 0.7$ in order to close the
scattering triangle). Points were weighted according to the number
of equivalent positions in the grey area of
Fig.~\ref{Fig:geo_fig.eps}(b).

\section{Results}
\subsection{Energy- and Wavevector-Dependent Scans}
\label{Sec:qw_scans}
\begin{figure}
\begin{center}
\includegraphics[width=0.8\linewidth]{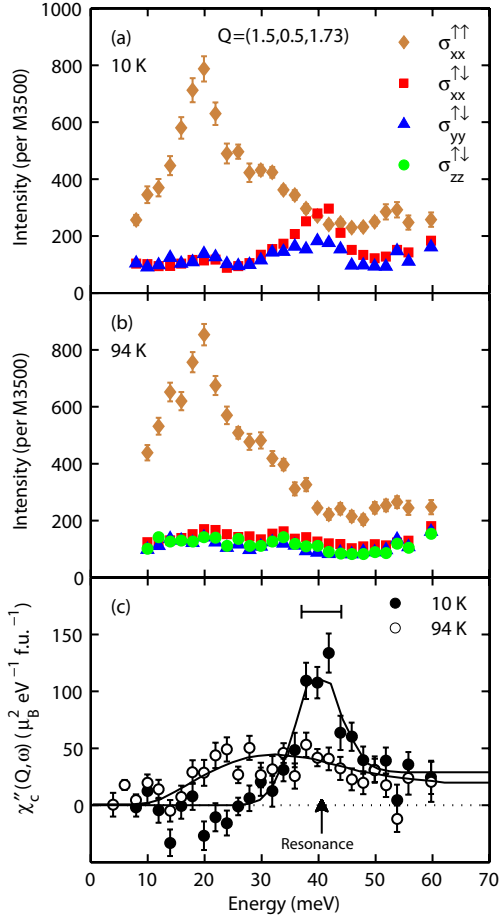}
\end{center}
\caption{(Color online) Energy-dependent scans with polarization
analysis collected at $\mathbf{Q}$=(1.5,0.5,1.73). (a-b) Spin-flip
and non-spin-slip cross sections for various spin polarizations in
the superconducting ($T=10$~K) and normal ($T=94$~K
$=T_c+1$~K) states. (c) Out-of-plane generalized susceptibility
$\chi_{c}$ determined from (a) and (b). The solid line for
the $T=10$~K data is a resolution corrected fit to the cross-section
described in the text. The horizontal bar represents the
full-width-at-half maximum (FWHM) resolution for a
$\delta(\omega-\omega_0)$ cross section.}
\label{Fig:YBCO9_pp_E_raw_chi_ab}
\end{figure}

\begin{figure}
\begin{center}
\includegraphics[width=0.95\linewidth]{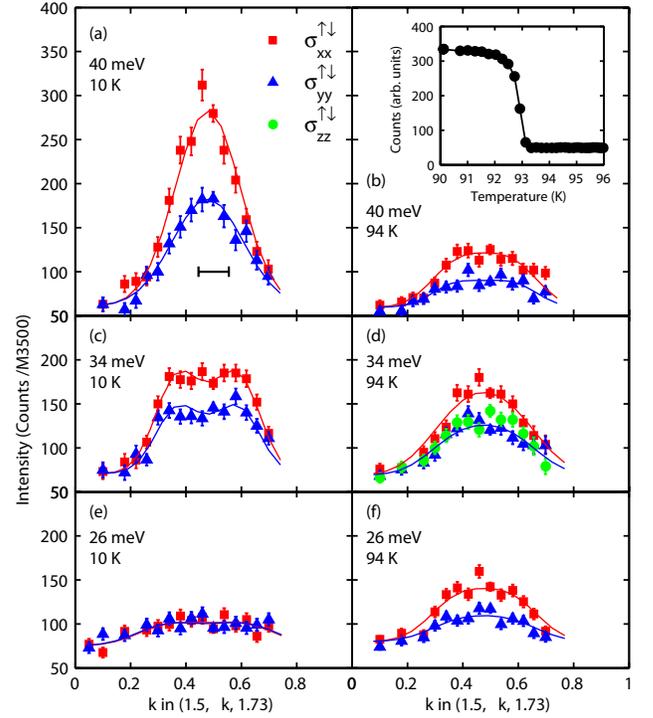}
\end{center}
\caption{(Color online) (a)-(f) Wavevector-dependent scans with LPA
through the (1.5,0.5,1.73) position at the resonance energy (a,b)
and lower energies (c-f). The solid lines are resolution corrected
fits to Eq.~\ref{Eqn:4peaks}. The horizontal bar represents the FWHM
resolution for a $\delta(\mathbf{Q}-\mathbf{Q}_0)$ cross-section.
The inset to (b) shows $T_c$ determined by a neutron depolarization
technique in which the sample was field cooled through $T_c$ in a
vertical field. The field was then rotated to be horizontal and the
spin flip scattering on the (310) Bragg peak measured on warming.}
\label{Fig:YBCO9_q_raw}
\end{figure}
\begin{figure}
\begin{center}
\includegraphics[width=0.95\linewidth]{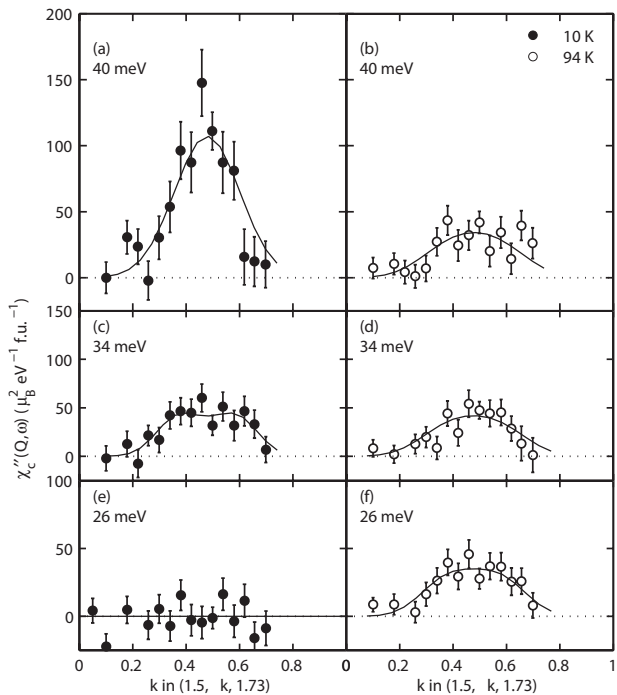}
\end{center}
\caption{The out-of-plane response $\chi_{c}^{\prime\prime}$ in the
normal and superconducting states determined from the data in
Fig.~\ref{Fig:YBCO9_q_raw}. The solid lines are resolution corrected
fits to the cross-section described in the text.}
\label{Fig:YBCO9_q_chi}
\end{figure}
\begin{figure*}
\begin{center}
\includegraphics[width=0.95\linewidth]{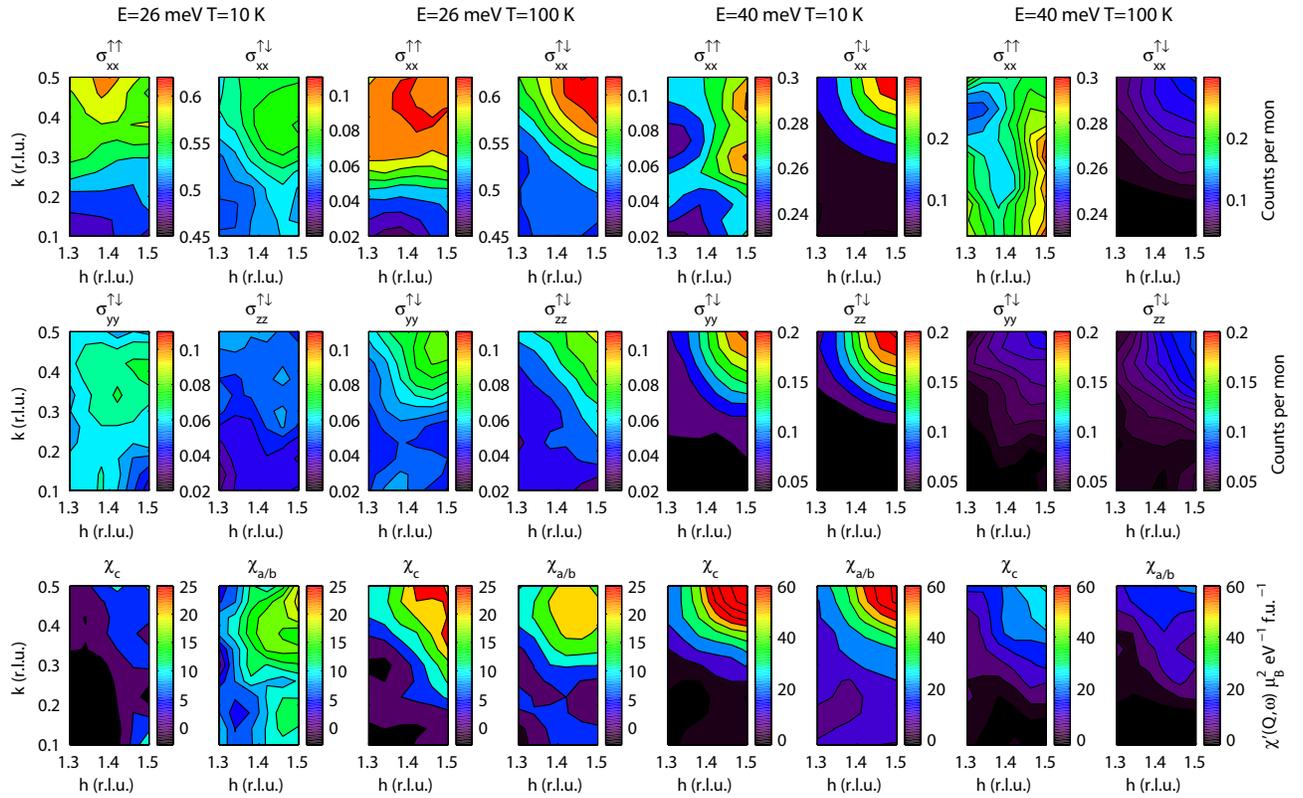}
\end{center}
\caption{(Color online) (Top two rows) Spin polarized cross sections
as defined in the text for $E$=26, 40~meV and $T$=10, 100~K. The
$\uparrow\uparrow$ channel is mostly phonon scattering and the
$\uparrow\downarrow$ channels are mostly magnetic scattering.
(Bottom row) Magnetic responses
$\chi_{a/b}^{\prime\prime}(\mathbf{q},\omega)$ and
$\chi_{c}^{\prime\prime}(\mathbf{q},\omega)$ determined from top two
rows. Note that some structure is due to statistical noise.}
\label{Fig:YBCO9_maps}
\end{figure*}
\begin{figure}
\begin{center}
\includegraphics[width=0.95\linewidth]{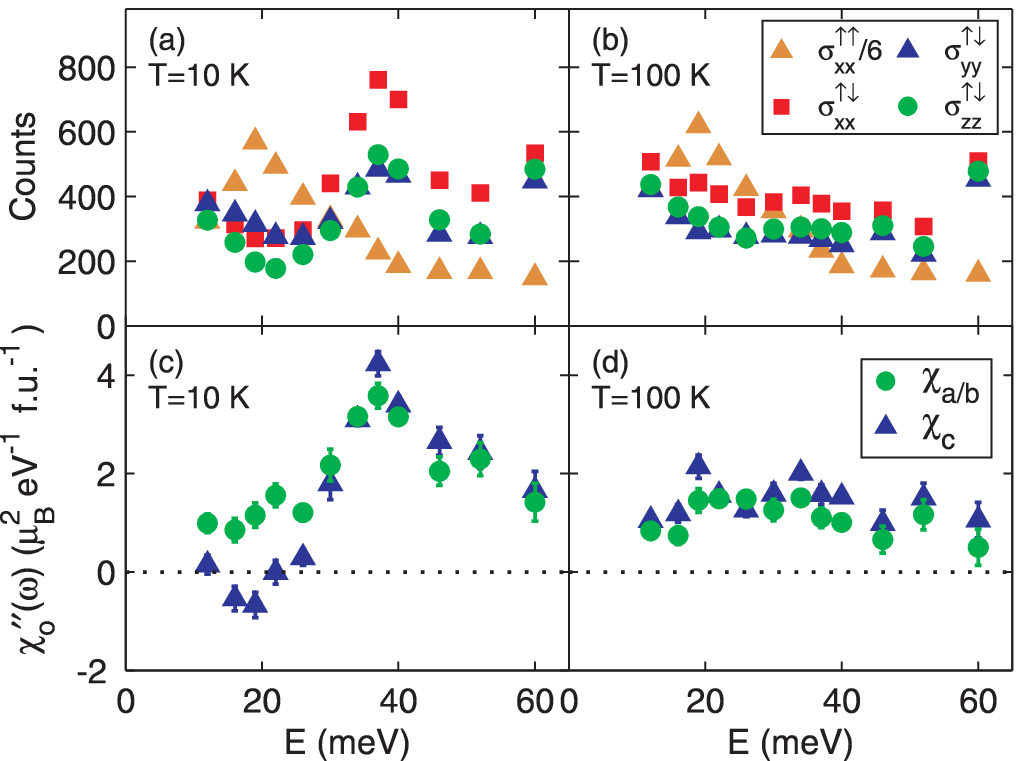}
\end{center}
\caption{(Color online) (a)-(b) Integral of the spin-polarized cross
sections over reciprocal space described in text and
Fig.~\ref{Fig:geo_fig.eps}. (c)-(d) Local susceptibility determined
from data in (a)-(b) via Eqs.~\ref{Eq:sigma_diff} and
\ref{Eqn:chi_local}.} \label{Fig:YBCO9_chi_local}
\end{figure}

Fig.~\ref{Fig:YBCO9_pp_E_raw_chi_ab} shows energy-dependent scans
made at the (1.5,0.5,1.73) position with various spin polarizations.
At this position in reciprocal space the non spin-flip (phonon)
scattering is up to 8 times stronger than the spin-flip scattering.
Thus an unpolarized measurement made under the same conditions would
be dominated by phonon scattering at some energies (the comparison with unpolarized experiments
is discussed further in Appendix~\ref{Sec:unpolarized}).  In the normal
state the $\sigma_{xx}^{\uparrow\downarrow}$ cross section is larger
than $\sigma_{yy}^{\uparrow\downarrow}$ and
$\sigma_{zz}^{\uparrow\downarrow}$ over a wide energy range, $20
\lesssim E \lesssim 60$~meV, signalling the presence of magnetic
excitations. We can use Eq.~\ref{Eqn:sigma2abc} to isolate the
out-of-plane response $\chi^{\prime\prime}_{c}$, this is shown in
Fig.~\ref{Fig:YBCO9_pp_E_raw_chi_ab}(c).  In the superconducting
state there is a large increase in
$\sigma_{xx}^{\uparrow\downarrow}$ and
$\sigma_{yy}^{\uparrow\downarrow}$
($\sigma_{zz}^{\uparrow\downarrow}$ was not measured in this case)
near the resonance energy. The difference scan
Fig.~\ref{Fig:YBCO9_pp_E_raw_chi_ab}(c) shows a sharp resonance peak
at $E \approx 40$~meV which appears to have formed by a transfer of
spectral weight from lower energies $E \lesssim 35$~meV. The
$\chi_{c}^{\prime\prime}$ response appears to be largely gapped
below about 30~meV.  Similar data was obtained using unpolarized
neutrons by Bourges~\textit{et al.} \cite{Bourges1999a}. We do not observe a collective magnetic excitation in the 50--60~meV range as observed recently in HgBa$_2$CuO$_{4+\delta}$ \cite{Li2010a}. We note that there is a peak in the non spin flip channel in this energy range in Fig.~\ref{Fig:YBCO9_pp_E_raw_chi_ab}(a).

In order to
analyze our data further, we fitted the $T=10$~K scan in
Fig.~\ref{Fig:YBCO9_pp_E_raw_chi_ab}(c) to the resolution-corrected
model cross section
\begin{equation}
\begin{split}
\chi_{c}\qw &=  \left[ A \delta(\omega-\omega_0) + B\theta(\omega-\omega_0) \right]  \\
& \times  \exp\left\{-\frac{(h-1/2)^2+(k-1/2)^2}{2\sigma^2}
\right\},
\end{split}
\label{Eqn:resonace}
\end{equation}
where $\theta$ is the heaviside step function and $\sigma$ is the
width parameter extracted from a $\mathbf{q}$-dependent scan through
the resonance (see Table~\ref{Tab:widths}). Throughout this paper we
use the RESTRAX simulation package \cite{Saroun1997a} to perform
convolutions of the instrumental resolution function and model cross
sections. Using the cross section defined by Eq.~\ref{Eqn:resonace},
we find that the width of the peak due to the resonance in
Fig.~\ref{Fig:YBCO9_pp_E_raw_chi_ab}(c) is resolution limited and
$\hbar\omega_0 = 41 \pm 1$~meV.

We have converted the data in
Fig.~\ref{Fig:YBCO9_pp_E_raw_chi_ab}(c) to absolute units using
Eq.~\ref{Eqn:cross_section} without attempting to deconvolve the
experimental resolution. This means that each point in the scan is
an average (in wavevector and energy) over the instrumental
resolution. Keeping this in mind, we have integrated the response in
Fig.~\ref{Fig:YBCO9_pp_E_raw_chi_ab}(c) in energy for $4<E<60$~meV
for $T$=10~K and 94~K. From Eqs.~\ref{Eqn:FD_theorem} and
\ref{Eqn:total_moment}, we find the out-of-plane fluctuating moments
$\langle m^2_{c} \rangle$ are 0.50$\pm$0.05 and 0.48$\pm$0.05
$\mu_B^2$f.u.$^{-1}$ at $T$=10~K and 94~K respectively (these are averaged over the resolution width in wavevector shown in Fig. \ref{Fig:YBCO9_q_raw}). Thus this
increase in the response at the resonance energy can be accounted
for by a shift in spectral weight from lower energies.

Fig.~\ref{Fig:YBCO9_q_raw} shows wavevector dependent scans along
the $(1.5,k,1.73)$ line at three characteristic energies.
Fig.~\ref{Fig:YBCO9_q_chi} shows the susceptibility extracted from
the data in Fig.~\ref{Fig:YBCO9_q_raw} using
Eq.~\ref{Eq:sigma_diff}. In the normal state ($T=94$~K), we observe
a magnetic response at all three energies. On cooling to $T=10$~K,
the lower frequency $E=26$~meV response is suppressed while the
response at the resonance energy ($E=40$~meV) increases dramatically
and the $\mathbf{q}$-width decreases. The data were fitted to a
model consisting of four incommensurate peaks with locations
$\mathbf{Q}_{\delta}=(1/2 \pm\delta,1/2)$ and $(1/2,1/2 \pm\delta)$
and width $\sigma$:
\begin{equation}
\chi^{\prime\prime}\qw=A\sum_{\mathbf{Q}_{\scriptstyle \delta}} \exp
\left\{ -\frac{(\mathbf{Q}-\mathbf{Q}_\delta)^2}{2 \sigma^2}
\right\}. \label{Eqn:4peaks}
\end{equation}
The results of this fitting procedure are shown in
Table~\ref{Tab:widths}
\begin{table}
\begin{tabular}{llcc}
  \hline
  $T$(K) & $\hbar\omega$(meV) & $\delta$(r.l.u) & $\sigma$(r.l.u) \\
  \hline
  10 & 26 & N/A  & N/A\\
     & 34 & $0.12\pm0.02$ & $0.059\pm0.01$ \\
     & 40 & $0$           & $0.114\pm0.01$ \\
  94 & 26 & $0.12$        & $0.085\pm0.01$ \\
     & 34 & $0.12$        & $0.095\pm0.01$ \\
     & 40 & $0.12$        & $0.071\pm0.01$ \\
     & 40 & $0$           & $0.16 \pm0.02$ \\
  \hline
\end{tabular}
\caption{ Incommensurability $\delta$ and width $\sigma$ parameters
obtained from fitting Eq.~\ref{Eqn:4peaks} to the scans in
Fig.~\ref{Fig:YBCO9_q_raw}. Where no error is quoted, the parameter
was fixed. \label{Tab:widths}}
\end{table}

We first consider the scans at the resonance energy
($\hbar\omega$=40~meV). A single Gaussian peak ($\delta$=0) provides
a good description of the scan in the superconducting state
[Fig.~\ref{Fig:YBCO9_q_raw}(a) and Fig.~\ref{Fig:YBCO9_q_chi}(a)].
In the normal state, there is magnetic scattering at the resonance
energy [Fig.~\ref{Fig:YBCO9_q_chi}(b)]. The existence of a magnetic
response at this energy in optimally doped YBCO has been a subject
of some debate
\cite{Rossat1991a,Mook1993a,Fong1995a,Bourges1999a,Dai2001a} and we
will discuss this later.  It is clear from our data that the
response at the resonance energy is broader in $\mathbf{q}$ and
weaker in the normal state than the superconducting state. If we fit
the 40~meV data using Eq.~\ref{Eqn:4peaks} with $\delta$=0, we find
$\sigma$=0.18$\pm$0.02 and 0.115$\pm$0.01 for the normal and
superconducting states respectively. Returning to the
superconducting state data at lower energy, we find a single
Gaussian peak ($\delta$=0) does not provide a good description of
the $\hbar\omega$=34~meV ($T$=10~K) scans
[Fig.~\ref{Fig:YBCO9_q_raw}(c) and Fig.~\ref{Fig:YBCO9_q_chi}(c)] in
the superconducting state. Better fits are obtained when a finite
incommensurability $\delta$=0.12$\pm$0.02 is used. This $\delta$ is
in agreement with that obtained in other studies of optimally doped
YBCO \cite{Dai2001a,Woo2006a}.  In the normal state
[Fig.~\ref{Fig:YBCO9_q_raw}(b,d,f) and
Fig.~\ref{Fig:YBCO9_q_chi}(b,d,f)] we see clear magnetic scattering
at the three energies investigated. We do not see a two-peaked
structure as in Fig.~\ref{Fig:YBCO9_q_raw}(c), instead the response
appears to be broadened out into single peak which, in some cases
[e.g. Fig.~\ref{Fig:YBCO9_q_raw}(b,f)], looks ``flat topped''. To
contrast the normal and superconducting state responses, we have
fitted the scans with the value of $\delta$ determined from the
$T$=10~K and $\hbar\omega$=34~meV scan. The normal state response is
broader in all cases (see Table~\ref{Tab:widths}).

\subsection{Local Susceptibility Measurements}
\label{Sec:local} In order to search for the diffuse contributions
to the magnetic response, we sampled a grid of points near the
(3/2,1/2) position where the response is generally stronger.
Extended grids at two characteristic energies are shown in
Fig.~\ref{Fig:YBCO9_maps}. For this part of the experiment we
collected three spin-flip channels and we were able to extract
$\chi_{a/b}^{\prime\prime}$ and $\chi_{c}^{\prime\prime}$. The
lowest row of Fig.~\ref{Fig:YBCO9_maps} shows the signal extracted
via Eq.~\ref{Eq:sigma_diff}. The data collected at $E=40$~meV shows
that the response is strongest near the (1.5,0.5,1.73) position both in
the normal and superconducting states.  At $E=26$~meV, we see a
normal state response which is spread out: see, for example,
$\chi_{a/b}^{\prime\prime}(E=26\mbox{ meV},T=100~\mbox{K})$, where the
upper part of the map shows signal.  On entering the superconducting
state $\chi_{c}^{\prime\prime}$ shows a much larger change than
$\chi_{a/b}^{\prime\prime}$ suggesting that a spin anisotropy
develops.  

Fig.~\ref{Fig:YBCO9_chi_local} shows the wavevector
integrals collected at a number of energies over the grey region
shown in Fig.~\ref{Fig:geo_fig.eps}. This is the region of highest
intensity in the Brillouin zone, but there is clearly scattering in
other parts of the zone. The contribution of the grey region to
$\chi_{o}^{\prime\prime}(\omega)$ is shown in
Fig.~\ref{Fig:YBCO9_chi_local}(c) and (d).
Fig.~\ref{Fig:YBCO9_chi_local} shows that there is a strong response
in the \textit{normal} state over a wide energy range.  When compared to the
energy-dependent scan at (1.5,0.5,1.73), we see that the higher energy
response is relatively stronger.  This is due to the presence of a
broader response in $\mathbf{q}$ at higher energies $E \gtrsim
50$~meV \cite{Hayden2004a,Reznik2004a,Woo2006a}. On entering the
\textit{superconducting} state, we see a strong reduction in
$\chi_{c}^{\prime\prime}$ with little change in
$\chi_{a/b}^{\prime\prime}$. This means the magnetic response develops
a strong spin anisotropy in the superconducting state (see
Sec.~\ref{Sec:Dis_aniso} for more discussion). For higher energies,
$E \geq 34$~meV, the response increases in the superconducting
state, not only at the resonance energy, but up to 60~meV.  Table~\ref{Tab:fluc_mom} shows that when integrated over the range $12<E<60$~meV the total fluctuating moment $\langle m^2 \rangle$ increases by about 60\%.  In order to compare with other studies of the resonance in near optimally doped YBCO \cite{Dai1999a,Fong1999a,Woo2006a}, we have also integrated the data in Fig.~\ref{Fig:YBCO9_chi_local} over the smaller energy range $30<E<60$~meV (see Table~\ref{Tab:fluc_mom}) in this case we see a larger change in $\langle m^2 \rangle$ (between the normal and superconducting states) which is comparable to previous reports \cite{Dai1999a,Fong1999a,Woo2006a}.

\begin{table}

\begin{tabular}{lccc}
  \hline
  $T$(K) & $\langle m^2_{a/b} \rangle(\mu_B^2 \; \mathrm{f.u.}^{-1})$ & $\langle m^2_{c} \rangle (\mu_B^2 \; \mathrm{f.u.}^{-1})$ & $\langle m^2 \rangle (\mu_B^2 \; \mathrm{f.u.}^{-1})$\\
  \hline
  \multicolumn{4}{c}{$12 \leq \hbar\omega(\mathrm{meV}) \leq 60 $} \\
  10     & $0.031\pm0.004$ &  $0.026\pm0.004$ & $0.088\pm0.007$\\
  100    & $0.017\pm0.003$ &  $0.022\pm0.003$ & $0.056\pm0.005$\\
    \hline
  \multicolumn{4}{c}{$30 \leq \hbar\omega(\mathrm{meV}) \leq 60 $} \\
  10     & $0.024\pm0.003$ & $0.026\pm0.003$  & $0.074\pm0.005$\\
  100    & $0.009\pm0.002$ & $0.014\pm0.002$  & $0.032\pm0.003$\\
  \hline
\end{tabular}
\caption{Fluctuating moments $\langle m^2_{a/b} \rangle$, $\langle
m^2_{c} \rangle$ and $\langle m^2\rangle=2\langle m^2_{a/b} \rangle +
\langle m^2_{c} \rangle$ in the normal ($T=100$~K) and superconducting ($T=10$~K)
states calculated by numerically integrating the response in
Fig.~\ref{Fig:YBCO9_chi_local}.
The errors quoted are statistical and do not include the systematic
error in the absolute scale which is about $\pm 20$\%.
\label{Tab:fluc_mom}}
\end{table}

\section{Discussion}
\begin{figure}
\begin{center}
\includegraphics[width=0.70\linewidth]{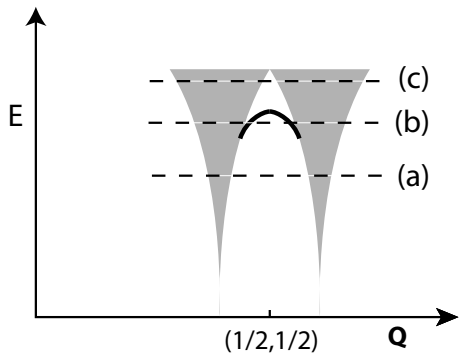}
\end{center}
\caption{Schematic representation of
$\chi^{\prime\prime}(\mathbf{q},\omega)$ in the superconducting
state of YBa$_2$Cu$_3$O$_{6.9}$ based on Refs.
\onlinecite{Onufrieva2002a,Eremin2005a}.  The black line is the
resonance mode and grey area the particle-hole continuum. Scans (a),
(b) and (c) correspond approximately to 20, 40 and 60 meV.}
\label{Fig:res_sketch}
\end{figure}

\subsection{Response in the Normal and Superconducting States}

Theories of the magnetic excitations in the superconducting state of
cuprate superconductors such as YBa$_2$Cu$_3$O$_{6+x}$ are well
developed
\cite{Lavagna1994a,Demler1995a,Liu1995a,Mazin1995a,Abanov1999a,Brinckmann1999a,Kao2000a,Norman2001a,
Tchernyshyov2001a,Onufrieva2002a,Eremin2005a,Eschrig2006a}. Many
features are explained by a magnetic exciton scenario
\cite{Liu1995a,Mazin1995a,Tchernyshyov2001a,Onufrieva2002a,Eremin2005a}
in which the resonance is a bound state in the particle-hole
channel, which appears in a region of $\mathbf{q}-\omega$ space
where there are no damping processes due to electron-hole pair
creation. This is illustrated schematically in
Fig.~\ref{Fig:res_sketch}.  In such a picture, significant magnetic
response should also be present in the normal state. As the system
enters the superconducting state we expect the low energy response
to be suppressed below $E\lesssim \Delta$ and an enhancement of the
response at the resonance energy. This is the behaviour seen in
Figs.~\ref{Fig:YBCO9_pp_E_raw_chi_ab} and \ref{Fig:YBCO9_q_chi}. The
nature of the magnetic response in the normal state of optimally
doped YBCO has been a subject of debate, particularly with regard to
energies near the resonance energy
\cite{Rossat1991a,Mook1993a,Fong1995a,Fong1996a,Bourges1999a,Dai2001a}.
Some studies suggest there is a significant response
\cite{Rossat1991a, Bourges1999a} for $\mathbf{q} \approx (1/2,1/2)$
and $\hbar\omega \approx 40$~meV, while others claim the response is
absent or too weak to observe \cite{Fong1995a,Fong1996a,Dai2001a}.
The present experiment allows the magnetic response to be separated
from phonon scattering. We find that the out-of-plane response
$\chi^{\prime\prime}_{c}(\mathbf{q},\omega)$ is peaked around
$\hbar\omega \approx 30$~meV for $\mathbf{q} \approx (1/2,1/2)$ in
the normal state ($T=94$~K). On cooling there is a shift of spectral
weight to higher energies which leads to the formation of the
resonance peak near 40 meV, with the concomitant formation of
incommensurate peaks observed at 34~meV and a spin gap below about
30~meV for the $\chi^{\prime\prime}_{c}$ component of the
response. This is consistent with the formation of a magnetic
excitonic mode as illustrated schematically in
Fig.~\ref{Fig:res_sketch}. The work presented in this paper refers
to optimally doped YBCO where it is harder to separate the magnetic
contribution from phonons and other background scattering than for
other compositions. We note that for underdoped YBCO (e.g.
YBa$_{2}$Cu$_{3}$O$_{6.6}$)
\cite{Mook1998a,Bourges2000a,Hayden2004a,Hinkov2007a} a strong
dispersive excitonic mode is also observed in the superconducting
state. On warming to $T_c$ the remnants of this mode are clearly
observable and persist well above $T_c$.

The discussion above relates to the energy- and wavevector-
dependent scans presented in Sec.~\ref{Sec:qw_scans}. These yield
information about the out-of-plane fluctuations described by
$\chi^{\prime\prime}_{c}$. We did not collect the corresponding
scans for $\chi^{\prime\prime}_{a/b}$, however, we did probe this
component of the local susceptibility in the measurements presented
in Sec.~\ref{Sec:local}.  These measurements were designed to yield
estimates for the total response in a region of $\mathbf{q}$ space
rather than identify the location of specific features such as
incommensurate peaks. They are summarized in
Fig.~\ref{Fig:YBCO9_chi_local}(c) and (d). In
Fig.~\ref{Fig:YBCO9_chi_local}(c) we see that there is strong
evidence for additional scattering below 30 meV in the
$\chi^{\prime\prime}_{a/b}$ component of the response. This response
appears to be rather spread out in wavevector when we inspect the
corresponding map ($\hbar\omega=$26~meV, $T=10$~K) in
Fig.~\ref{Fig:YBCO9_maps}.  Thus our results suggest that there are
other (diffuse) contributions to the $\chi^{\prime\prime}_{a/b}$
response at low energies in the superconducting state.  The
$\chi^{\prime\prime}_{a/b}$ component of the response has a lower
`spin gap' than the $\chi^{\prime\prime}_{c}$ component. The low
energy response ($E \lesssim 30$~meV) may be due to the
electron-hole continuum also present in theories of the resonance
\cite{Tchernyshyov2001a,Onufrieva2002a,Eremin2005a}. This is
illustrated schematically in Fig.~\ref{Fig:res_sketch}.

\subsection{Spin Anisotropy in YBa$_2$Cu$_3$O$_{6.9}$}
\label{Sec:Dis_aniso} Our results suggest that a spin anisotropy
develops in the lower energy ($10$$\lesssim$$E$$\lesssim$30~meV) excitations on
entering the superconducting state.  
Nuclear magnetic resonance (NMR) probes the spin fluctuations in the
very low frequency limit and, indeed, the \textit{anisotropy} of
spin-lattice relaxation rate ($T_1$) in YBa$_2$Cu$_3$O$_{7}$ has
been reported to show a strong temperature dependence in the
superconducting state \cite{Barrett1991a,Takigawa1991a}. 
Various theories have attributed this to the combined effect of the NMR form
factor and a changing $\chi^{\prime\prime}(\mathbf{q},\omega)$ (See e.g. Ref.~\onlinecite{Bulut1992a,Thelen1993a}). 
However, the present measurements show that there is also an significant intrinsic anisotropy in $\chi_{\alpha}^{\prime\prime}(\mathbf{q},\omega)$ with respect to the spin direction which must be considered. 
It is interesting to note that Uldry \textit{et al.} \cite{Uldry2005a} have extracted the intrinsic anisotropy from NMR data and concluded that the out-of-plane correlations do not change appreciably on entering the superconducting state, in contrast to our results. This may be because NMR measurements probe the excitations at much lower frequencies than our measurements.   

Anisotropy in the susceptibility ultimately comes from the
spin-orbit interaction. An exotic case is the superfluid $^3$He
A-phase \cite{Vollhardt1990a}, where the susceptibility depends on
the orientation of the angle of the field to the characteristic spin
vector $\mathbf{d}$. In the case of superconductors, dramatic
changes in a pre-existing spin anisotropy have recently been
observed in BaFe$_{1.9}$Ni$_{0.1}$As$_2$ \cite{Lipscombe2010a} and a small anisotropy at the resonance energy is observed in FeSe$_{0.5}$Te$_{0.5}$ \cite{Babkevich2010a}. A possible origin of the emergence of spin anisotropy in
YBa$_2$Cu$_3$O$_{6.9}$ may be the Dzyaloshinskii-Moriya (DM)
interactions between the copper spins \cite{Coffey1991a}. The
buckled structure of the CuO$_2$ planes in ortho-I
YBa$_2$Cu$_3$O$_{6.9}$ (see Fig.~\ref{Fig:YBCO_struct}) means that
DM interactions of the form $\mathbf{D} \cdot \mathbf{S}_i \times
\mathbf{S}_j$ are allowed between neighbouring Cu spins. The
presence of such terms leads to additional spin anisotropy. This
leads to a polarization dependence to the spin wave dispersion and
energy in the antiferromagnetic parent compounds La$_{2}$CuO$_{4}$
\cite{Peters1988a} and YBa$_{2}$Cu$_{3}$O$_{6.2}$
\cite{Shamoto1993a}. In the case of YBa$_{2}$Cu$_{3}$O$_{6.2}$ the
anisotropy gaps are $\sim$10~meV \cite{Shamoto1993a} and the ordered
moments lies along the [100] direction \cite{Janossy1999a}.

The low energy excitations ($E\lesssim$30~meV) we observe have their predominant
fluctuations within the CuO$_2$ planes making the $a/b$ response largest.  At higher energies, $E \approx$40~meV, the
excitations are more isotopic. This corresponds to all three components
of the spin-triplet $\{ |\uparrow \uparrow \rangle, |
\uparrow\downarrow\rangle- \downarrow\uparrow\rangle, |\downarrow
\downarrow \rangle \}$ being excited.

\section{Conclusion}
In this work we used inelastic neutron scattering with longitudinal
polarization analysis to measure the magnetic excitations in the
normal and superconducting states of near optimally doped
YBa$_{2}$Cu$_{3}$O$_{6.9}$.  We have unambiguously identified a
strong magnetic response in the \textit{normal state} which appears to exist
over the 10--60~meV range of the present experiment.  On entering
the \textit{superconducting state}, the out-of-plane magnetic response
($\chi_{c}^{\prime\prime}$), is strongly suppressed at lower
energies, while the response at the magnetic resonance energy and
above increases.  We also find evidence for a new diffuse component
to the magnetic response in the  $\chi_{a/b}^{\prime\prime}$ component
of the susceptibility at low energies 10$\lesssim$$E$$\lesssim$30~meV which is present in the superconducting state.

\section{Acknowledgements}
We would like to acknowledge helpful discussion with James Annett,
Anthony  Carrington, PengCheng Dai, Chris Lester, Jan \v{S}aroun, Nic Shannon and
Qimiao Si.

\appendix
\section{Sum Rules and the Magnetic Response}

\subsection{Local Susceptibility}
\label{Sec:chi_local} The local susceptibility is a useful way to
characterise the overall response.  It is defined as,
\begin{equation}
\label{Eqn:chi_local} \chi^{\prime\prime}(\omega)= \frac{\int
\chi^{\prime\prime}(\mathbf{Q},\omega) \; d^3\mathbf{Q} } {\int
d^3\mathbf{Q}},
\end{equation}
where, in general, the integrals are over a volume of reciprocal
space which samples the full $\mathbf{Q}$ dependence of
$\chi^{\prime\prime}(\mathbf{Q},\omega)$. In the case of
YBa$_{2}$Cu$_{3}$O$_{6+x}$ this is one unit cell in the $ab$ plane
and infinity along $c$. The local susceptibility can be split into
the two terms of Eq.~\ref{Eqn:chi_oe_def}. Thus integrating
Eq.~\ref{Eqn:chi_oe_def} we have
\begin{equation}
\chi^{\prime\prime}(\omega)=\chi_o^{\prime\prime}(\omega)+\chi_e^{\prime\prime}(\omega),
\end{equation}
where
\begin{equation}
\chi_o^{\prime\prime}(\omega)=\frac{1}{2} \int_{0}^{1} \; dh
\int_{0}^{1} dk \; \chi_o^{\prime\prime}(h,k,\omega).
\end{equation}
The definition for $\chi_o^{\prime\prime}(\omega)$ used here differs
by a factor 2 from earlier work, but allows a direct comparison with
single layer compounds \cite{Vignolle2007a}.

\subsection{Total Moment Sum Rule}
For an ion with spin only moment, the total squared moment is
\begin{eqnarray}
\nonumber \langle m^2 \rangle & = & g^2 \mu^2_B S(S+1) \\
& = & 3 \mu_B^2 \mbox{ for } S=\frac{1}{2} \mbox{ and } g=2.
\end{eqnarray}
The total fluctuating moment observed by INS over a given range of
energy and momentum can be determined from the
fluctuation-dissipation theorem and is
\begin{eqnarray}
\nonumber \langle m^2 \rangle & = & \langle m_x^2+m_y^2+m_z^2  \rangle \\
& = & \frac{1}{\pi} \int \left[
\frac{\chi_{xx}^{\prime\prime}(\omega)+\chi_{yy}^{\prime\prime}(\omega)+\chi_{zz}^{\prime\prime}(\omega)}
{1-\exp(-\hbar \omega/kT)} \right] \; d\omega.
\label{Eqn:total_moment}
\end{eqnarray}

\section{Comparison with Unpolarized Studies}
\label{Sec:unpolarized}
\begin{figure}
\begin{center}
\includegraphics[width=0.8\linewidth]{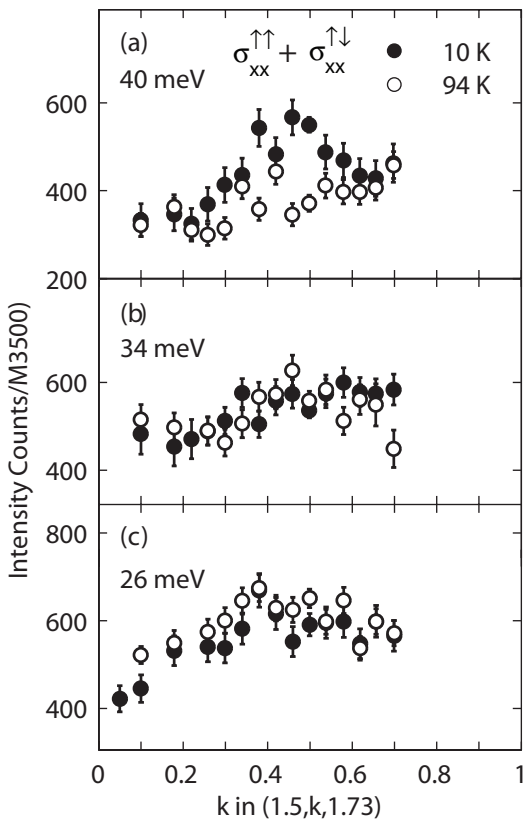}
\end{center}
\caption{Wavevector-dependent scans of $\sigma_{xx}^{\uparrow\uparrow}+\sigma_{xx}^{\uparrow\downarrow}$ at various energies. This combination allows comparison with unpolarized studies.  \label{Fig:Qscan_unpol}}
\end{figure}

\begin{figure}
\begin{center}
\includegraphics[width=0.8\linewidth]{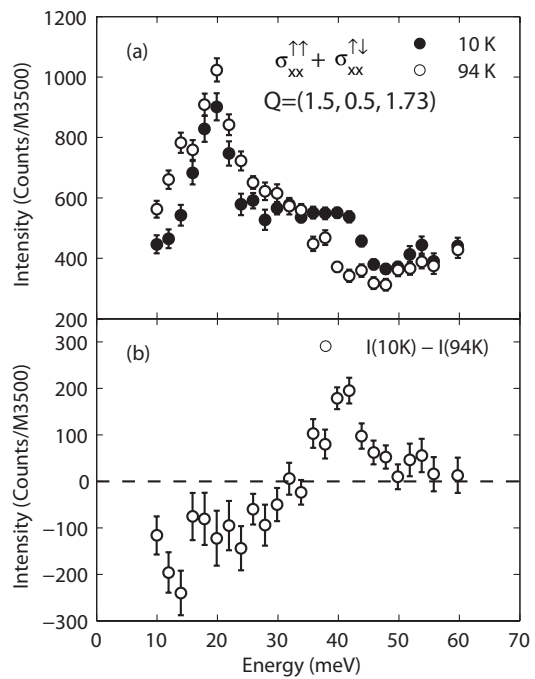}
\end{center}
\caption{(a) Energy-dependent scans of $\sigma_{xx}^{\uparrow\uparrow}+\sigma_{xx}^{\uparrow\downarrow}$ at $\mathbf{Q}$=(1.5,0.5,1.73) at $T=10,94$~K. (b) Difference of scans in (a). \label{Fig:Escan_unpol}}
\end{figure}

There are many \textit{unpolarized} studies of the magnetic excitations in YBa$_{2}$Cu$_{3}$O$_{6+x}$ \cite{Rossat1991a,Fong1995a,Stock2005a,Dai2001a,Reznik2004a}. In this section we show that our results are broadly consistent with previous results.  The main issues that arise in unpolarized studies are: (i) the separation of magnetic signal from background and (ii) the separation of magnetic and phonon scattering. In the present spin-polarized study we may compare to different spin-flip cross-sections to remove the background and the phonon contribution. This is demonstrated in Eqs.~\ref{Eqn:all_sigmas}-\ref{Eq:sigma_diff}.

The unpolarized inelastic cross section is generally of the form
\begin{equation}
\left( \frac{d^2 \sigma}{d \Omega dE} \right) \propto
\frac{\chi^{\prime\prime}(\mathbf{q},\omega,T)}{1-\exp(-\hbar \omega/k T)}+N(q,\omega,T),
\label{Eqn:sigma_unpolarized}
\end{equation}
where the first term represents the inelastic magnetic response and the second that due to the phonons. A sharp magnetic response such as the resonance can be isolated through $\mathbf{q}$ and $\omega$ scans and verified as being magnetic through the form factor present in Eq.~\ref{Eqn:cross_section}.  However, a broad or diffuse response is more difficult to distinguish from phonons. The phonon response $N(q,\omega,T)$ usually decreases with temperature ($\hbar \omega \lesssim k T$) or remains constant ($\hbar \omega \gg k T$) due to the Bose factor. Thus a signal that increases with decreasing temperature (such as the resonance) is likely to be magnetic. If a magnetic signal decreases with decreasing temperature] such as the response below about 30~meV in Fig.~\ref{Fig:YBCO9_pp_E_raw_chi_ab}(c)] it is difficult to distinguish from phonons using unpolarized neutrons.

In Figs.~\ref{Fig:Qscan_unpol} and \ref{Fig:Escan_unpol}, we have reconstructed `unpolarized' scans by adding together the spin-flip and non-spin-flip intensities for $\mathbf{H} \parallel x$, $\sigma_{xx}^{\uparrow\uparrow}+\sigma_{xx}^{\uparrow\downarrow}$.  Our experiment was not optimized for this reconstruction because the spin-flip channels were counted longer than non-spin-flip, nevertheless we can make some useful observations. As expected, Fig.~\ref{Fig:Qscan_unpol}(a)  clearly shows the resonance at $T=10$~K and $E=40$~meV in the superconducting state. Note there is increased background or phonon scattering at larger $k$ in this scan. In the normal state, at $T=94$~K, it is not possible to identify any magnetic scattering. For $E=34$~meV [Fig.~\ref{Fig:Qscan_unpol}(b)], the scans at both temperatures are similar. The data are consistent with a broad magnetic response which changes little between the two temperatures [see Fig.~\ref{Fig:YBCO9_q_chi}(c)-(d)]. Finally, for $E=26$~meV we observe a decrease in intensity across much of the scan on lowering the temperature. This is consistent with a reduction of the magnetic response at this energy [see Fig.~\ref{Fig:YBCO9_q_chi}(e)-(f)]. However, the phonon scattering at this energy and wavevector is strong [see Fig.~\ref{Fig:YBCO9_pp_E_raw_chi_ab}(a)] thus part (about 50\%) the reduction observed using unpolarized spectroscopy is due to the change of the Bose factor for the phonons.

Fig.~\ref{Fig:Escan_unpol} shows  energy-dependent scans at the $\mathbf{Q}$=(1.5,0.5,1.73) position and a temperature difference often used to isolate the resonance (see e.g. \cite{Fong1995a,Fong1999a}). From Figs.~\ref{Fig:YBCO9_pp_E_raw_chi_ab} and \ref{Fig:YBCO9_q_raw}, we can deduce that about 50\% of the observed change observed with temperature at 26~meV is due to phonons.


%

\end{document}